\documentclass[preprint,12pt]{elsarticle}

\usepackage{amssymb}
\usepackage{amsmath}

\journal{Nuclear Physics B}

\newtheorem{definition}{Definition}
\usepackage{hyperref}
\usepackage{xurl}
\usepackage{multirow}
\usepackage{graphicx}
\usepackage[table,xcdraw]{xcolor}
\usepackage{booktabs}
\usepackage{makecell}         
\usepackage{subfigure}

\usepackage[toc, acronym]{glossaries}
\makeglossaries

\newacronym{sg}{SG}{Synchronous Generator}
\newacronym{ibr}{IBR}{Inverter-Based Resource}
\newacronym{TSO}{TSO}{Transmission System Operator}
\newacronym{WAMS}{WAMS}{Wide-Area Monitoring System} 
\newacronym{PMU}{PMU}{Phasor Measurement Unit}
\newacronym{ML}{ML}{Machine Learning}
\newacronym{GAN}{GAN}{Generative Adversarial Network}
\newacronym{LHS}{LHS}{Latin Hypercube Sampling}
\newacronym{OP}{OP}{operating point}
\newacronym{gfol}{GFL}{grid-following}
\newacronym{gfor}{GFM}{grid-forming}
\newacronym{PF}{PF}{power flow}
\newacronym{OPF}{OPF}{optimal power flow}
\newacronym{HPC}{HPC}{high-performance computing}
\newacronym{PLL}{PLL}{Phase-Locked Loop}
\newacronym{XGB}{XGB}{Extreme Gradient Boosting}

\usepackage{hyperref}


\begin{document}

\begin{frontmatter}



\title{Data Generation for Stability Studies of Power Systems with High Penetration of Inverter-Based Resources
}

\author[label1]{Francesca~Rossi\corref{cor1}}
\author[label2]{Mauro~Garcia~Lorenzo}
\author[label2]{Eduardo~Iraola~de~Acevedo}
\author[label1]{Elia~Mateu~Barriendos}
\author[label1]{Vinicius~Albernaz~Lacerda}
\author[label2]{Francesc~Lordan\textendash Gomis}
\author[label2]{Rosa~Badia}
\author[label1]{Eduardo~Prieto\textendash Araujo}

\affiliation[label1]{organization={Department of Electrical Engineering, CITCEA-UPC},
            addressline={Avinguda Diagonal, 647}, 
            city={Barcelona},
            postcode={08028}, 
            state={Barcelona},
            country={Spain}}

\affiliation[label2]{organization={Barcelona Supercomputing Center},
            addressline={Plaça d'Eusebi Güell, 1-3, Les Corts}, 
            city={Barcelona},
            postcode={08034}, 
            state={Catalonia},
            country={Spain}}
            
\cortext[cor1]{Corresponding author e-mail: francesca.rossi@upc.edu}
\begin{abstract}
The increasing penetration of inverter-based resources (IBRs) is fundamentally reshaping power system dynamics and creating new challenges for stability assessment. Data-driven approaches, and in particular machine learning models, require large and representative datasets that capture how system stability varies across a wide range of operating conditions and control settings. This paper presents an open-source, high-performance computing framework for the systematic generation of such datasets. The proposed tool defines a scalable operating space for large-scale power systems, explores it through an adaptive sampling strategy guided by sensitivity analysis, and performs small-signal stability assessments to populate a high-information-content dataset. The framework efficiently targets regions near the stability margin while maintaining broad coverage of feasible operating conditions. The workflow is fully implemented in Python and designed for parallel execution. The resulting tool enables the creation of high-quality datasets that support data-driven stability studies in modern power systems with high IBR penetration.
\end{abstract}




\end{frontmatter}


\printglossaries   

\section{Introduction}
The ongoing energy transition involves the progressive phase-out of fossil-fuel-based \glspl{sg}, the large-scale integration of renewable \glspl{ibr}, and the increasing electrification of demand. As a consequence, power systems are evolving into power-electronics-dominated systems. In these systems, electromagnetic phenomena become increasingly relevant in shaping the overall dynamic behavior, complementing and in some cases influencing more strongly the traditional electromechanical dynamics that characterize grids dominated by \glspl{sg}~\cite{hatziargyriou2020definition}. These phenomena appear on distinct time scales: electromechanical dynamics evolve relatively slowly, typically over seconds, whereas electromagnetic dynamics occur on the order of microseconds due to the high switching frequencies and fast control loops of power electronic converters~\cite{hatziargyriou2020definition}. As a consequence, modern electrical networks are increasingly exposed to fast and potentially poorly damped oscillations that require rapid detection and mitigation. Predicting these events in advance is particularly challenging because of the intermittent and volatile nature of renewable energy sources.
To address these new operational challenges, \glspl{TSO} are acting both on the monitoring side and on the stability detection and assessment side. On the monitoring side, they are equipping their networks with advanced real-time measurement infrastructures such as \glspl{PMU} and \glspl{WAMS}~\cite{CIGRE2023_WAMPAC,SEGUNDOSEVILLA2022107772}. On the assessment side, there is growing interest in \gls{ML}-based surrogate models as alternatives to conventional time-domain simulations~\cite{xu2020intelligent}. Conventional simulations provide high accuracy but are computationally expensive. \gls{ML}-based models, instead, offer the possibility of delivering fast stability assessments suitable for real-time or near–real-time operation~\cite{xu2020intelligent}.

High-quality training data are essential for the development of \gls{ML} models for power system stability assessment. Relying solely on historical records from system operation, such as available \gls{PMU} measurements and corresponding stability labels, is not sufficient to train accurate models, because these data are not fully representative of the operating conditions of interest~\cite{xu2020intelligent,thams2019efficient}. Power systems are traditionally operated in a conservative manner, which means that undamped oscillations and instability events are rare. Although the few events that do occur are valuable for understanding the mechanisms that trigger them, they are not sufficient to derive statistically meaningful patterns. \gls{ML} classifiers, such as those used to predict stability status, require balanced datasets~\cite{5128907} and therefore a significant number of unstable \glspl{OP}. In addition, high-quality information in the region of the operating space close to the stability boundary is critical, and it has been shown to be the most important factor for obtaining accurate models~\cite{giraud2025dataset}. 

For these reasons, historical data alone cannot provide an adequate training dataset. They must be enriched with synthetic instances generated through simulation. The sampling strategy used to generate such synthetic data should be guided by three main principles: historical relevance, sufficient coverage of the operating space, and discriminative relevance to capture rare events and operating points near the stability margin~\cite{bugaje2023generating}. Considering historical relevance, this aspect can be addressed by sampling the operating space according to historical consumption and generation patterns. The study in~\cite{gillioz2025large} follows this idea by generating long-term synthetic time series derived from historical national data, producing realistic bus-level power injections for \gls{ML} applications. To further enhance the representativeness of historically based samples, generative models and other data-driven techniques have been proposed. In~\cite{lan2024data}, the authors build on historical steady-state data by using \gls{GAN}-based augmentation and transfer learning to recreate rare operating conditions that are underrepresented in real records. In~\cite{xia2024database}, historical variability is incorporated by modeling the probability distribution of operating states under uncertain loads and renewable production, generating samples in proportion to their likelihood of occurrence. Although these methods significantly improve realism, they still do not achieve sufficient coverage of the full operating space.

To improve coverage, random sampling techniques such as Monte Carlo sampling and \gls{LHS} are typically employed~\cite{duchesne2020recent}. Using random sampling, however, introduce a major challenge: a large number of infeasible \glspl{OP} is produced, i.e. points for which the \gls{PF} does not converge or for which the solution cannot be used in real operation because it does not satisfy static stability constraints, such as maintaining voltages within acceptable ranges, or because it violates operational limits of generators and transmission lines. The issue of feasibility in sampled \glspl{OP} is addressed in the studies in~\cite{venzke2021efficient,bugaje2023generating}. In~\cite{venzke2021efficient}, the authors formulate infeasibility certificates based on separating hyperplanes to classify regions of the operating space as infeasible and exclude them from sampling. This removes large portions of the search space and improves scalability, but may also eliminate feasible OPs. The remaining unclassified space is then explored using directed walks to identify the stability margin, followed by sampling from a multivariate distribution fitted to the secure points. Although directed walks initiated from multiple starting points can be parallelized, the overall workflow still follows a fixed sequence of steps: space reduction, directed walks, and sampling in the stability margin. In~\cite{bugaje2023split,bugaje2023generating}, a power-system adaptation of the split-based GAPSPLIT strategy is proposed. Although modifications are introduced to prevent the algorithm from getting trapped in infeasible regions, the method fundamentally relies on recursively bisecting the input domain. As a result, it remains fully sequential and cannot be parallelized. Moreover, existing methods do not incorporate dimensionality reduction and instead operate in a high-dimensional space (or polytope~\cite{venzke2021efficient}), whose dimensionality grows exponentially with system size. Finally, these approaches do not investigate the impact of \glspl{ibr}, whose presence can significantly jeopardize system stability.

The limitations of existing sampling approaches highlight the need for data generation frameworks that explicitly account for the dynamic behavior of systems with high levels of \glspl{ibr}, enhance scalability by reducing the effective dimensionality of the operating space, and exploit advances in computational technologies such as \gls{HPC} and parallelization. This paper addresses these gaps through the following main contributions:
\begin{itemize}
    \item A data generation framework tailored to power systems with high \glspl{ibr} penetration is introduced. The tool is designed to capture how system dynamics evolve under varying penetration levels of \gls{gfor} and \gls{gfol} \glspl{ibr}, as well as under different controller tuning settings.

    \item A systematic methodology for operating-space design is presented, addressing the challenge of high dimensionality by defining how relevant dimensions are selected and structured, thereby enhancing scalability while preserving system complexity.

    \item An efficient exploration strategy for operating-space sampling is developed. This strategy incorporates mechanisms to identify and prioritize regions with high information content, specifically those near the stability margin. It also includes an online sensitivity analysis that identifies the most influential dimensions and steers adaptive sampling accordingly.

    \item A workflow optimized for \gls{HPC} environments is provided, enabling effective parallelization and efficient execution of large-scale data generation and stability analysis tasks.

    \item A plug-and-play, open-source tool is released, fully implemented in Python and available at https://github.com/MauroGarciaLorenzo/datagen.
\end{itemize}
The remainder of the paper is organized as follows. Section~\ref{sec: data_gen} presents the proposed data generation tool. Section~\ref{sec: case_study} applies the tool to a case study and evaluates its performance under different parameter settings. Finally, conclusions are drawn in Section~\ref{sec: conclusion}.

\section{Data Generation Tool}
\label{sec: data_gen}
\begin{figure}
    \centering
    \includegraphics[width=0.5\linewidth]{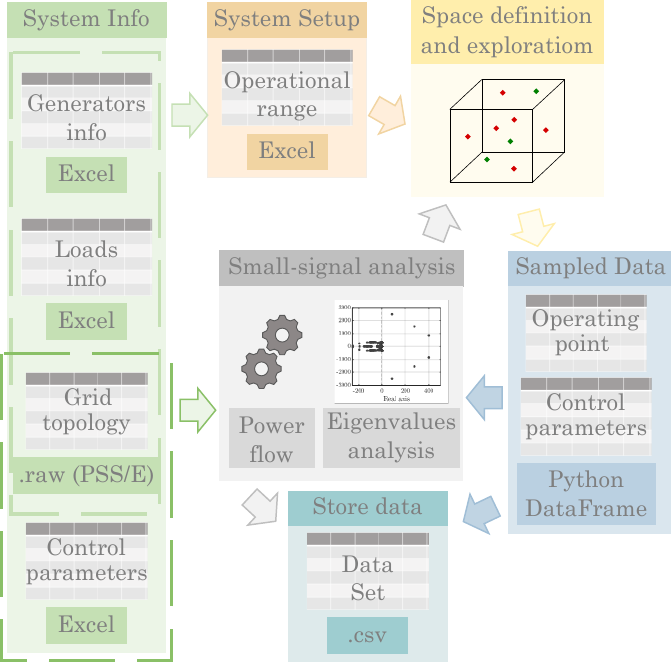}
    \caption{Data generation tool workflow.}
    \label{fig:PF_SS}
\end{figure}

This section describes the proposed data generation tool, including the input information it requires, how this information is processed, the workflow of the data generation procedure, and the characteristics of the resulting dataset. The main inputs, outputs, and processing steps are summarized in Fig.\ref{fig:PF_SS}. The required inputs include the system information for which the data must be generated, such as generation data, demand profiles, and the grid topology. These inputs can be provided in tabular form (as Excel files), while the network topology can also be supplied through standard RAW files. A detailed description of the input quantities is provided in Section~\ref{sec:sys_setup}. Based on these inputs, the system's range of operation is identified and translated into the operating space to be explored. To ensure a scalable exploration of this space, the methodology used to define its dimensions is presented in Section~\ref{sec:dims}. The operating space is then explored through an iterative data generation process. In each iteration, relevant quantities related to the operating points and controller settings are sampled, and a small-signal stability analysis is performed. Both the sampled variables and the quantities computed during the stability analysis are stored to build the final dataset, which can be exported as a tabular CSV file. The results of each iteration are also processed internally to drive the exploration toward the stability margin, thereby increasing the efficiency and information content of the generated dataset.

\subsection{System Setup for Dynamic Studies}
\label{sec:sys_setup}
Prior to initiating the data generation process, it is essential to establish the configuration of the system under analysis. This involves two key steps: (i) defining the system’s operational range and (ii) identifying an appropriate aggregation of generators for dynamic analysis. The methodology proposed to perform these tasks is detailed below.

\subsubsection{Range of Operation}
Determining the operational range is essential to define the boundaries of the operational space that the data generation process must explore. This step involves specifying the lower and upper operation limits for each individual generator and load within the system. The level of detail that can be achieved in this phase depends significantly on the availability of system information. Nonetheless, it is generally reasonable to assume that a minimum dataset includes the following elements: 
\paragraph{Generation-related information}
\begin{itemize}
\item The bus to which each generator is connected;
\item The installed capacity of each generator, expressed in terms of rated and nominal power;
\item The minimum and maximum active and reactive power output, defined either by applicable grid codes or by the generator’s specific capability curve;
\item The generation technology type (e.g., fossil-based, nuclear, hydroelectric, photovoltaic, wind), to determine whether the unit should be modeled as a \gls{sg} or an \gls{ibr}.
\end{itemize}
If the analysis aims to assess the impact of different asset control schemes and settings, it is assumed that the control strategies and configurations to be tested are known. For each control parameter intended to be varied, a reasonable range is then established based on grid codes or control design principles.
\paragraph{Load-related information}
\begin{itemize}
\item The bus to which each load is connected;
\item Historical load profiles, either disaggregated by individual loads or aggregated by system regions;
\item The share of total system demand attributed to each load.
\end{itemize}

The aforementioned information is processed and organized into a structured table, which serves as input to the data generation tool for modeling the operational space to be explored. Specifically, on the generation side, the table includes, for each bus: the total nominal power installed; the maximum and minimum active and reactive power that can be injected or absorbed; and the same set of parameters disaggregated by generator type, i.e., for the total \gls{sg} and \gls{ibr} present at the bus.

\subsubsection{Equivalents for dynamic studies}
To conduct dynamic studies of bulk power systems, it is necessary to simplify the system modeling. In particular, representing each generator individually is impractical; therefore, generator aggregation must be employed. A basic aggregation approach may group generators based on their technology type, such as \glspl{sg} or \glspl{ibr}. However, to account for the varying control strategies applied to \glspl{ibr} (primarily \gls{gfol} versus \gls{gfor}) a different aggregation approach can be implemented, distinguishing generators according to their control mode. To verify the validity of this control-based aggregation approach from a dynamic behavior perspective, an admittance-based frequency scan can be used as a verification method, as shown in Appendix~\ref{appendix: dyn_eq}. Specifically, the conducted frequency response scans prove that the dynamic behavior observed by the grid remains consistent whether all \glspl{ibr} connected to a bus are modeled individually or aggregated into a single equivalent converter, provided they share the same control strategy. In addition, they reveal that, for identical operating points, the system's dynamic response differs significantly depending on whether the \glspl{ibr} are controlled in \gls{gfor} or \gls{gfol} mode. This observation highlights the importance of accounting for control strategy when analyzing system stability. It further motivates the generation of a dataset that captures stability assessments across various operating points and different levels of \gls{gfol}/\gls{gfor} penetration, while maintaining the same overall operating conditions. 

\subsection{Definition of The Operating Space}
\label{sec:dims}
\begin{figure}[ht]
    \centering
    \includegraphics[width=0.5\linewidth]{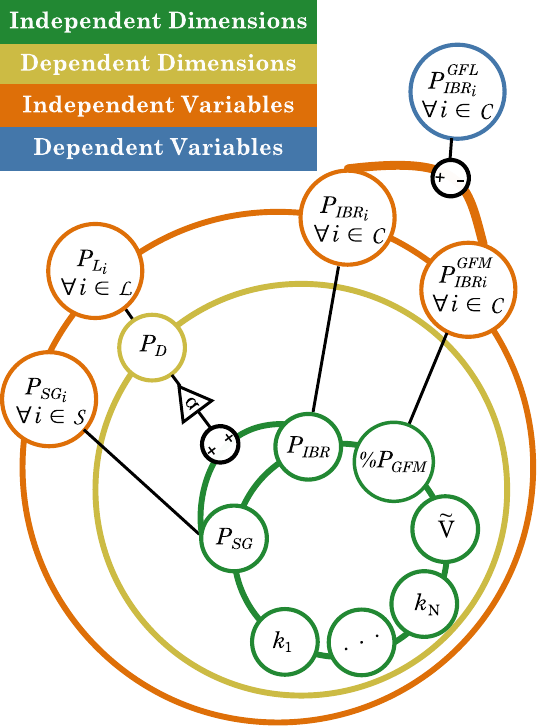}
    \caption{Dimensions and variables of the operating space.}
    \label{fig:tab_dim_var}
\end{figure}
Once the system’s operational range is identified, the corresponding operating space can be established and subsequently explored during the data generation process, following the proposed sampling strategy. The definition of the operating space to be explored has a critical impact on both the scalability and feasibility of the proposed data generation process. On one hand, the space must be sufficiently comprehensive to ensure that all quantities required for the stability assessment of each operating point can be appropriately sampled, in accordance with the input specifications of the tools employed in the stability analysis. On the other hand, to ensure scalability, it is not practical to define one dimension for every single quantity to be sampled. Instead, the operating space must be structured to meet the sampling requirements while minimizing the number of dimensions. In terms of feasibility, it is essential to account for the underlying relationships between the variables to be sampled. Ignoring these relationships and applying purely random sampling across all variables would result in a large number of infeasible operating points, making the exploration of the space unmanageable. To address these challenges, the operating space is represented using a set of selected dimensions and associated variables. Below, we provide a formal definition of the identified types of dimensions and variables, along with a clear specification of the quantities they represent. The overall structure is summarized in Figure~\ref{fig:tab_dim_var}.

\begin{definition}{\bf{(Independent Dimensions)}}\label{def: dim}
Independent Dimensions are system-level variables that can be sampled independently. 
\end{definition}
They represent key quantities, such as total power generation by technology type, voltage profiles, and control settings, that can be freely varied to generate operating scenarios.
Specifically, on the generation side, the primary quantity to be sampled is the total power injected into the grid. This dimension can be decomposed into two separate components: the total active power injected by \glspl{sg} ($P_{SG}$) and \glspl{ibr} ($P_{IBR}$), whose sum equals the total system generation. Since generators are modeled as PV nodes in the power flow analysis, their voltage setpoints must also be sampled. This corresponds to including the grid voltage profile (\textit{\~{V}}) as an additional dimension of the operating space. On the control side, if multiple control schemes or modes are being analyzed, further dimensions must be introduced. These include: the share of power injected by \glspl{ibr} operating under a specific control strategy (e.g., \gls{gfor} ($\%P_{GFM}$)), and one independent variable for each controller parameter selected for variation ($k_1$,...,$k_n$).

\begin{definition}\textbf{(Dependent Dimension)} The dependent dimensions refer to system-level quantities whose value cannot be sampled independently, but it depends on those sampled for the independent dimensions.    
\end{definition}
A key example is the total power demand ($P_D$), which must be consistent with the total generation. To account for system losses, the total demand can be set as a fraction of the total generation, calculated as the sum of the sampled active power contributions from \glspl{sg} and \glspl{ibr}.
\begin{definition}\textbf{(Independent Variables)}
Variables are the quantities assigned to individual system elements (e.g., generators, loads) based on the values sampled from the corresponding independent or dependent dimensions. They disaggregate the aggregated quantities defined by the dimensions and are used to construct detailed operating points for simulation and analysis.
\end{definition}
The variables associated with the independent dimensions serve to distribute the sampled quantities across individual generators. Specifically, they determine the power to be injected by each \gls{sg} ($P_{SG_i} \forall i \in \mathcal{S}$, where $\mathcal{S}$ is the set of \glspl{sg}), by each \gls{ibr}, and by each \gls{ibr} operating under the same control strategy defined in the corresponding dimension ($P_{IBR_i}$ and $P_{GFM_i}$ $\forall i \in \mathcal{C}$, where $\mathcal{C}$ is the set of converters in the system).\\
Concerning the dependent dimension $P_D$, its associated variables are responsible for distributing 
the total load 
across individual loads ($P_{L_i} \forall i \in \mathcal{L}$, where $\mathcal{L}$ is the set of loads in the system).
\begin{definition}\textbf{(Dependent Variables)}\\
Dependent variables are quantities that result from disaggregating values sampled in the dimensions, whose assignment is determined by the values of other independent variables. They cannot be independently sampled and are instead derived to maintain consistency within the defined operating point.
\end{definition}

Specifically, they include the power injected by each \gls{ibr} that operates under a control strategy different from the one explicitly sampled (e.g., \gls{gfol} ($P_{GFL_i}$ $\forall i \in \mathcal{C}$)). These values are determined as the difference between the total \gls{ibr} power and the portion allocated to \glspl{ibr} operating in \gls{gfor} mode, computed on a generator-unit basis.


\subsection{Sampling Methods}
To further enhance the variability of the sampled instances, the sampling of dimensions and variables can be organized in a hierarchical manner. A predefined number of samples ($n_{\text{samples}}$) may first be generated for the dimensional quantities. For each of these dimensional samples, an additional ($n_{\text{cases}}$) number of samples may then be generated for the associated variables. For example, $n_{\text{samples}}$ values of the dimensional quantity $P_{\mathrm{SG}}$ can be drawn, and for each of these values generate $n_{\text{cases}}$ combinations of the individual generator variables $P_{\mathrm{SG},i}$. This allows the analysis of how stability is affected when the same total synchronous generation participation is maintained, but different set-point distributions are assigned across the generators. Building on these considerations, the following sampling methods are proposed for the generation of dimensional values and their corresponding variables.

\subsubsection{Independent Dimensions Sampling}
The sampling of dimension quantities aims to efficiently maximize coverage of the operating space. To this end, the \gls{LHS} technique is applied to all independent dimensions, except for the voltage profile ($\tilde{V}$). The voltage profile must be treated with a different approach compared to the other independent dimensions. Although the goal is to test the system’s behavior under conditions characterized by various voltage profiles, it is important to consider that the voltage at each node cannot be assigned completely at random, as the overall voltage profile of the network is determined by physical constraints and electrical laws governing its operation. Supporting this, it is well known that a network tends to exhibit a natural voltage profile that characterizes its normal operating regime. However, such information may not be available a priori. Therefore, to reconcile random sampling with the need to obtain realistic profiles that reflect the system’s physical behavior, in the absence of detailed knowledge of the conventional voltage profile, the following approach is proposed. A starting bus (e.g., the slack bus) is first selected and assigned a random voltage magnitude within the permissible operating range defined by the grid. The remaining network is then traversed as a graph, assigning to each neighboring bus a voltage magnitude computed as the voltage of the previously visited bus plus a random deviation bounded within a specified range. In the case of meshed networks, where a bus is likely to be reached through multiple paths, its final voltage value is taken as the average of the multiple tentative assignments.

\subsubsection{Independent Variables Sampling}
When sampling the values of the independent variables, different strategies can be employed depending on the available information and the intended goal: whether it is to maximize coverage of the variable space, focus on extreme values, or replicate historically observed patterns. The implemented sampling methods include an approach that maximizes variance by starting each case at the lower bounds and incrementally increasing variable values in a shuffled order, allocating the remaining sum until the total matches the target; if this method fails to converge within set limits, a fallback is used that samples from Gaussian distributions centered at scaled means for each variable, with standard deviations chosen to keep values within bounds, and then proportionally adjusts the results so their sum precisely matches the desired total.

\subsection{Space Exploration Workflow}
\begin{figure}[ht]
    \centering
    \includegraphics[width=\linewidth]{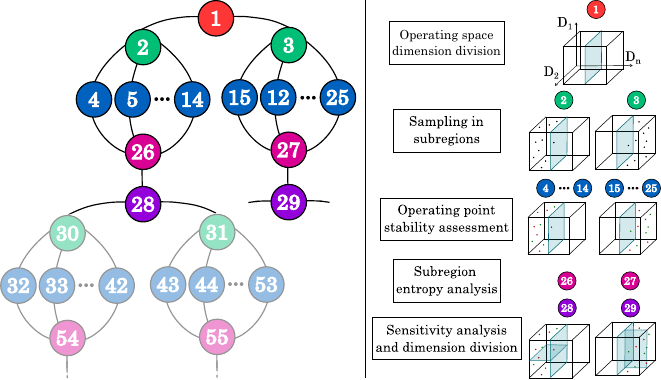}
    \caption{Definition of the exploration process.}
    \label{fig:exploration-workflow}
\end{figure}

The exploration begins with the definition of the operating space, followed by an iterative process that comprises five main steps: (i) dividing the space into subregions, (ii) sampling operating points within each subregion, (iii) assessing the stability of the sampled points, (iv) performing entropy analysis in each subregion, and (v) conducting sensitivity analysis to guide further space division. The exploratory algorithm is illustrated by Figure~\ref{fig:exploration-workflow}. This process continues until a predefined stopping criterion is met. 

The stability assessment is performed using conventional small-signal stability analysis tools, as described in Appendix~\ref{appendix: SS_FEASPF}. Specifically, this step involves:
\begin{itemize}
    \item Computing the feasible power flow solution (through an \gls{OPF}) for each sampled \gls{OP}, which allows determining \gls{OP} quantities not explicitly set during sampling (e.g., bus voltage phase angles) and adjusting the sampled quantities to values that ensure \gls{PF} feasibility;
    \item evaluating the stability of each operating point. 
\end{itemize}

The recursive subdivision of subregions aims to increase the exploration granularity, but only within areas of the space identified as of primary importance for the analysis, namely those intersected by the system’s stability margin. To enhance the efficiency of the exploration process, regions that do not contribute additional information are excluded from further sampling. This includes regions that are identified as infeasible (i.e., \gls{OPF} does not converge — see Appendix~\ref{appendix: SS_FEASPF}), or consist almost entirely of stable or unstable points. 

To determine whether a subregion should be further divided into child subregions, thereby continuing the exploration in that area, or whether the exploration in that subregion should instead be stopped, criteria based on the ratio of feasible points and on metrics that assess the diversity of stability behavior are applied. For the latter, an entropy-based metric is used, where entropy is computed over the binary stability classification (stable or unstable) of the samples contained in the subregion. Consider that a \textit{parent} subregion is subdivided into two \textit{child} subregions, each \textit{child} receives the samples that were previously generated within its corresponding portion of the space during the exploration of the \textit{parent}. The cutoff criteria that determine whether subdivision and further exploration should continue are defined as follows:
\begin{itemize}
    \item The entropy is zero, indicating that all sampled points within the subregion belong to a single class (either stable or unstable), and thus no stability margin is detected.
    \item The entropy decrease between two consecutive subdivisions falls below a predefined threshold, suggesting that further subdivision is unlikely to yield significantly new information.
    \item Minimum feasible rate, discontinuing exploration in areas where the proportion of infeasible cases is too high.
    \item A minimum tolerance is achieved, to stop dividing a dimension if its current range falls below a specified percentage of the initial range, to avoid exploring overly narrow regions.
    \item A maximum subdivision depth is reached, serving as a safeguard to limit computational effort.
\end{itemize}

The choice of which dimension to divide when a subregion is split can either be fixed in advance, by specifying the dimensions that should be explored with greater effort, or it can be guided by a sensitivity analysis that identifies the dimension along which the stability classification shows the highest variability. In this second case, a Random Forest is trained using the samples generated up to that point. The model provides an estimate of the most important features involved in the classification process~\cite{hastie2009elements}, and this information is used to select the dimensions that will be split in the next iteration of the workflow.

\subsection{HPC Application}
Parallelization is organized around two task types. The first is the subregion exploration task, which carries out the random sampling, entropy evaluation, grid subdivision, and recursive calls needed to analyze the resulting \textit{child} subregions. The second is the stability assessment task, which runs the ~\gls{OPF} and the small-signal stability analysis described in Appendix~\ref{appendix: SS_FEASPF}. As the exploration task calls itself and triggers the stability assessment task at each step, the parallelization framework must support recursive and nested execution. This is handled through distributed agents, each capable of launching a local runtime on its assigned node.
This produces a task dependency tree where a small number of exploration tasks form the upper structure, and a much larger number of stability assessment function calls branch below (one for each sampled \gls{OP}).


\section{Case Study}
\label{sec: case_study}
This section demonstrates the application of the proposed data generation tool to a power system representative of a highly \gls{ibr}-penetrated grid. The test system is first described, after which datasets are generated both with and without the use of the sensitivity-driven sampling strategy. The resulting datasets are then analyzed to assess the process used to generate them, their composition, and their quality through the performance of the \gls{ML} models trained on them. It is worth noting that omitting the sensitivity analysis results in a data generation process that resembles the approach proposed by the authors in earlier work~\cite{rossi2022data}. Thus, while the tool presented in this paper is inherently better suited for large systems and high-dimensional operating spaces, the case study is also designed to validate the added value of the sensitivity-based strategy, which represents one of the key novelties of this work.

The simulations are executed in a \gls{HPC} environment, specifically on the MareNostrum supercomputer at the Barcelona Supercomputing Center. Parallelization and distributed data exploration and generation are managed using PyCOMPSs~\cite{tejedor2017pycompss}, the Python interface of the COMPSs framework~\cite{badia2015comp}. PyCOMPSs relies on a distributed runtime composed of multiple agents that coordinate task scheduling and data management across the computing nodes~\cite{lordan2014servicess}, enabling transparent and efficient execution of parallel workflows in large-scale \gls{HPC} environments.

\subsection{System Setup}
\begin{figure}[ht]
    \centering
    \includegraphics[width=0.7\linewidth]{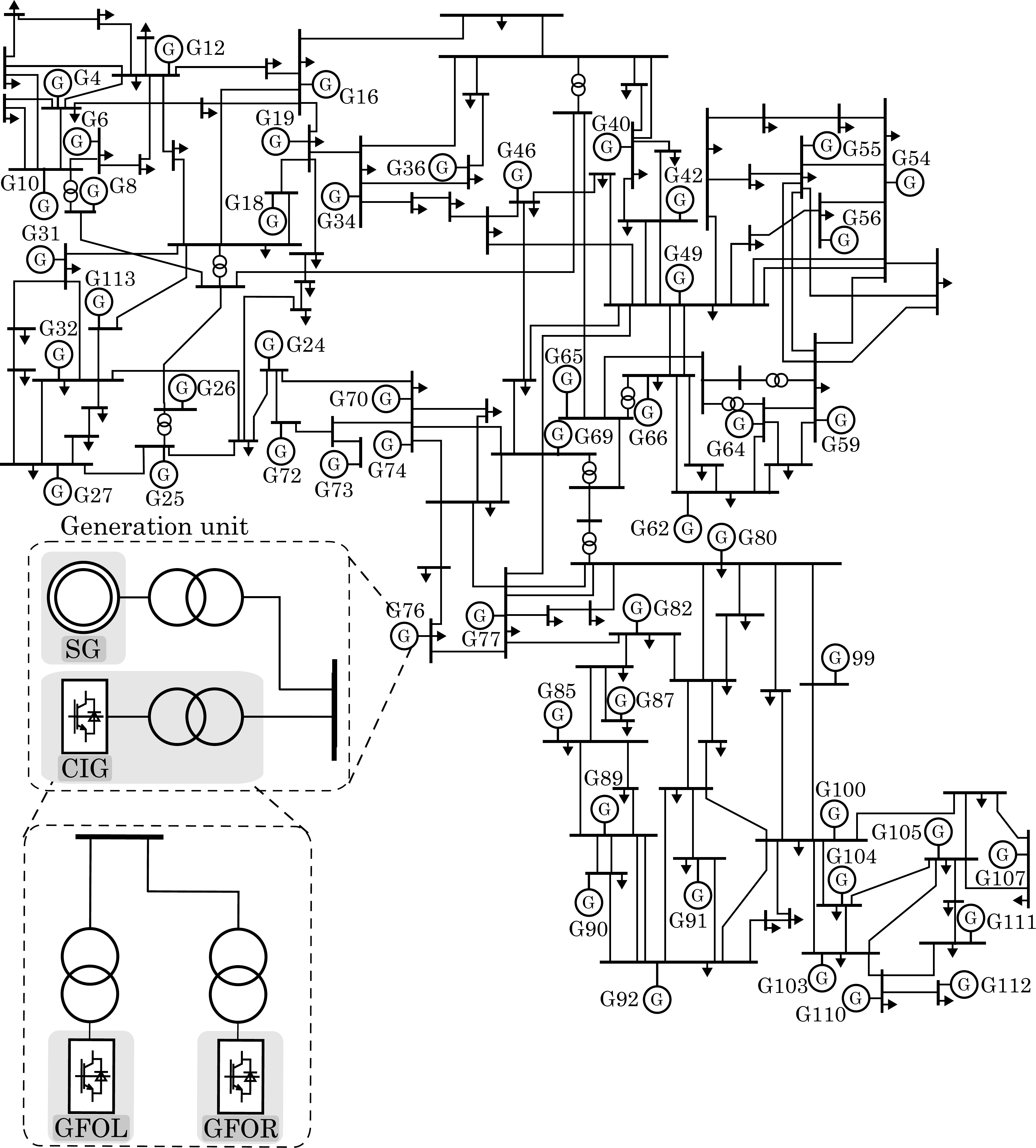}
    \caption{Scheme of the NREL 118-bus system.}
    \label{fig:nrel118}
\end{figure}

The system used to test and validate the data generation tool is the NREL 118-bus system, a modified version of the IEEE 118-bus system characterized by higher installed demand and generation capacity compared to earlier versions. The network is shown in Figure~\ref{fig:nrel118}. The following subsections present the main features of the system that are relevant for modeling purposes within the data generation framework. For a complete and detailed description, the reader is referred to~\cite{pena2017extended}, which also provides the full datasets describing the system.

\paragraph{Generation}
\renewcommand{\arraystretch}{1.5} 
\begin{table}[t]
\caption{Summary of the calculation procedure used to set the generation operational range.}
\label{tab:op_quant}
\centering
\begin{tabular}{ccccc}
\hline
\rowcolor[HTML]{D9D9D9} 
\multicolumn{1}{|c|}{\cellcolor[HTML]{D9D9D9}Bus} & 
\multicolumn{1}{c|}{\cellcolor[HTML]{D9D9D9}$P_{SG}^{\max}$} & 
\multicolumn{1}{c|}{\cellcolor[HTML]{D9D9D9}$P_{IBR}^{\max}$} & 
\multicolumn{1}{c|}{\cellcolor[HTML]{D9D9D9}$P^{\max}$} & \multicolumn{1}{c|}{\cellcolor[HTML]{D9D9D9}$S_{SG,rated}$} \\ \hline
\multicolumn{1}{|c|}{$n$} & 
\multicolumn{1}{c|}{= $P_{SG,nom}$} & 
\multicolumn{1}{c|}{= $P_{IBR,nom}$} & 
\multicolumn{1}{c|}{= $P_{SG}^{\max}+P_{IBR}^{max}$} & 
\multicolumn{1}{c|}{= $\frac{P_{SG,nom}}{\cos\varphi}$} \\
\hline
 & & & &\\
 \hline
\multicolumn{1}{|c|}{\cellcolor[HTML]{D9D9D9}$S_{IBR,rated}$} & 
\multicolumn{1}{c|}{\cellcolor[HTML]{D9D9D9}$S_{rated}$} & 
\multicolumn{1}{c|}{\cellcolor[HTML]{D9D9D9}$P_{SG}^{\min}$} & 
\multicolumn{1}{c|}{\cellcolor[HTML]{DCDADA}$P_{IBR}^{\min}$} & 
\multicolumn{1}{c|}{\cellcolor[HTML]{DCDADA}$P^{\min}$} 
\\ 
\hline
\multicolumn{1}{|c|}{= $\frac{P_{IBR,nom}}{\cos\varphi}$} & 
\multicolumn{1}{c|}{= $\frac{P_{nom}}{\cos\varphi}$} & 
\multicolumn{1}{l|}{=20\% of $S_{SG,rated}$} & \multicolumn{1}{l|}{=20\% of $S_{rated}^{IBR}$} & \multicolumn{1}{l|}{=20\% of $S_{rated}$}  \\ 
\hline
 &  &  &  &  \\ \cline{3-4}
 &  & \multicolumn{1}{|c|}{\cellcolor[HTML]{DCDADA}$Q^{\min}$} & 
\multicolumn{1}{c|}{\cellcolor[HTML]{DCDADA}$Q^{\max}$} \\ \cline{3-4}
 &  & 
\multicolumn{1}{|c|}{= - $S_{rated} \sin\varphi$} & 
\multicolumn{1}{c|}{= $S_{rated} \sin\varphi$} \\ \cline{3-4}
\end{tabular}
\end{table}
The NREL 118-bus system includes 24.6~GW of installed generation capacity and incorporates ten different power generation technologies. In total, there are 327 generators distributed across 54 generation units (i.e., generation buses), as illustrated in Figure~\ref{fig:nrel118}. Of these, 17 are wind power plants and 35 are photovoltaic solar plants. These renewable facilities represent the \glspl{ibr} in the system and account for 18.3\% of the total installed capacity. The remaining generation capacity is based on \gls{sg} technology and includes steam turbines powered by coal, gas, and other fuels; internal combustion engines powered by gas; combustion turbines powered by gas and oil; gas combined-cycle turbines; as well as hydro and biomass generators. The dataset provided in~\cite{pena2017extended} 
specifies, for each generator, its nominal power and the bus to which it is connected. From this information, generators are aggregated by type (\gls{sg}, \gls{ibr}, and total) to compute, at each bus, the total nominal and rated power as well as the minimum and maximum active and reactive power injections. The rated power and injection limits are derived from the generator capability curves; for simplicity, all generators are assumed to follow the same curve that sets the active power injection larger than 20\% of the rated power, and the reactive power injection with a power factor $\cos\varphi \geq$ 0.95. 
Table~\ref{tab:op_quant} summarizes the calculation procedure used to obtain each of the above quantities.

\paragraph{Demand}
The system comprises 91 loads. The dataset provided in~\cite{pena2017extended} 
specifies, for each load, its \emph{Load Participation Factor}, i.e., the fraction of total system demand it represents. 

\begin{figure}[ht]
\vspace{-0.3cm}
    \centering
    \includegraphics[width=0.8\linewidth]{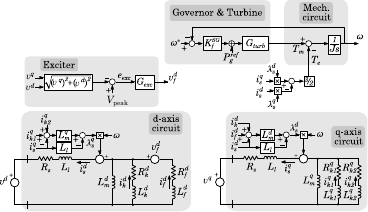}
    \caption{\gls{sg} control scheme.}
    \label{fig:sg_control}
\end{figure}

\begin{figure}[ht!]
    \centering
    \subfigure[\gls{gfol} converter control scheme.\label{fig:gfol_control_scheme}]{
        \centering
        \includegraphics[width=0.7\linewidth]{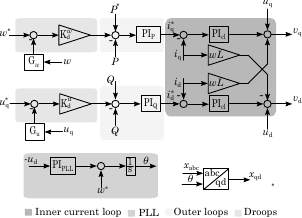}
        }
    \subfigure[\gls{gfor} converter control scheme.\label{fig:gfor_control_scheme}]{
        \centering
        \includegraphics[width=0.8\linewidth]{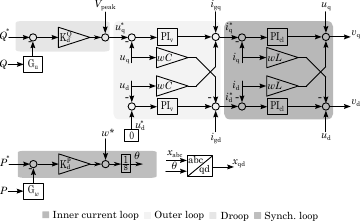}
        }   
    \caption{Converters control schemes.}
    \label{fig:control_schemes}
\end{figure}

\paragraph{SG and IBR models for small-signal analysis}
The \gls{sg} and \gls{ibr} models considered in this study are those shown in Figure~\ref{fig:sg_control}-\ref{fig:control_schemes}. A detailed description of the \gls{sg} model can be found in~\cite{8933043}. The \gls{sg} model includes the generator, with its electrical circuits represented in the $dq$ reference frame, and a single-mass model capturing the mechanical dynamics. It also incorporates an exciter
, a governor, and turbine models.

The \gls{gfol} converter control scheme includes an inner current control loop, a \gls{PLL}, two outer loops for active and reactive power control, and two droop controls: frequency droop and voltage droop. The \gls{gfor} converter control, instead, consists of an inner current control loop, outer voltage control loop, and two droop controls: an active power droop (used for synchronization) and a reactive power droop. In both \gls{gfol} and \gls{gfor} schemes, each droop controller is preceded by a first-order low-pass filter $G(s) = \frac{1}{\tau s + 1}$,
where $\tau$ is the filter time constant. The values of $\tau$ for all droop filters ($G_u$ and $G_w$ in Figures~\ref{fig:gfol_control_scheme}–\ref{fig:gfor_control_scheme}) are considered in this study as independent converter control parameters and are thus treated as control-related independent dimensions.

\subsection{Analysis of The Generated Datasets}
To validate the performance of the proposed tool, datasets are generated under different parameter settings. The resulting data are then analyzed in terms of their composition and how they evolve during the execution of the data generation process. Finally, an \gls{XGB} model~\cite{xgboost_python} is trained, and its accuracy is evaluated.

\begin{figure}
    \centering
    \includegraphics[width=0.7\linewidth]{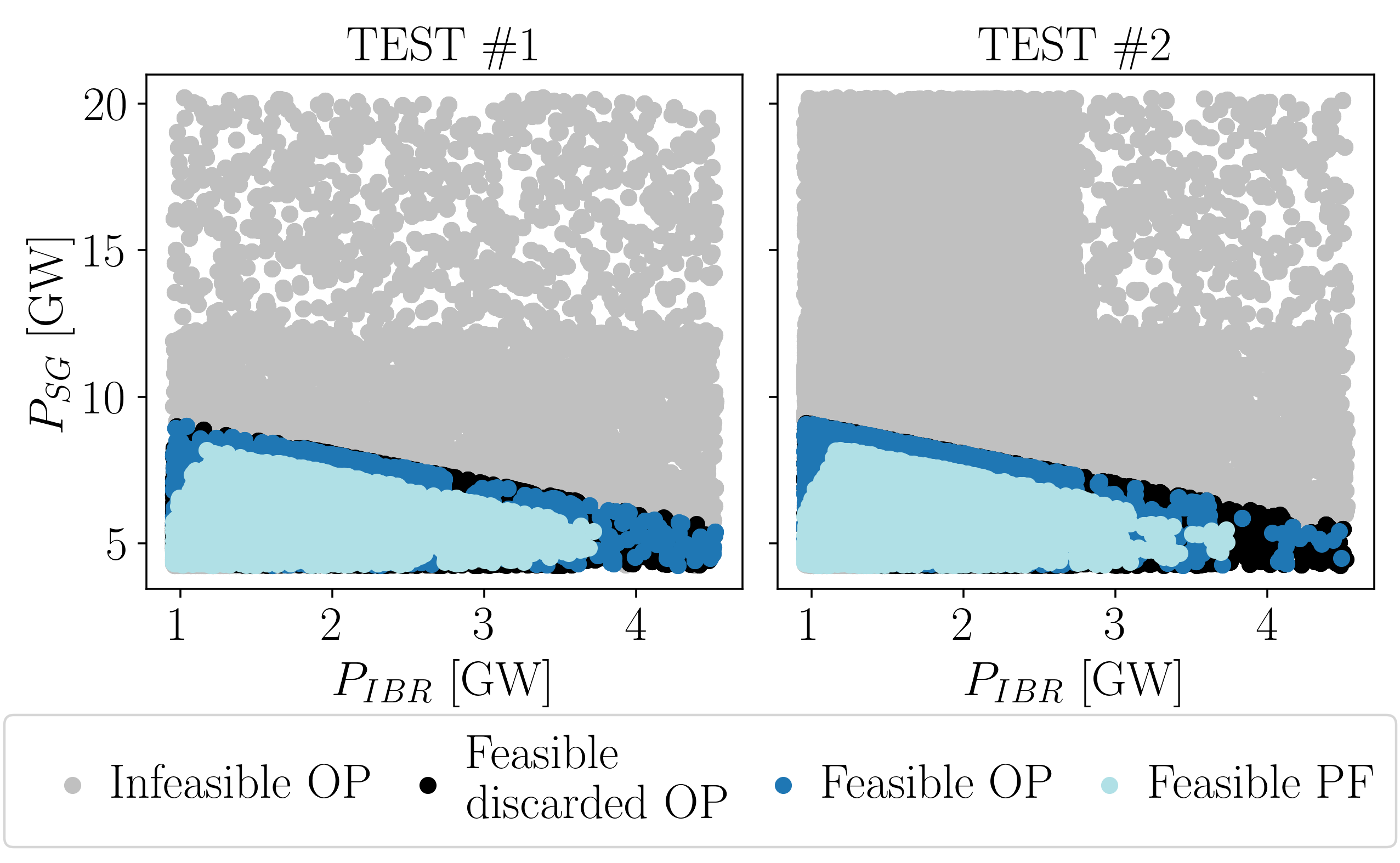}
    \caption{Feasible and infeasible \glspl{OP} and \gls{PF} solutions for the two test cases.}
    \label{fig:p_sg_p_cig_feasible_points}
\end{figure}

\begin{figure}
    \centering
    \includegraphics[width=0.7\linewidth]{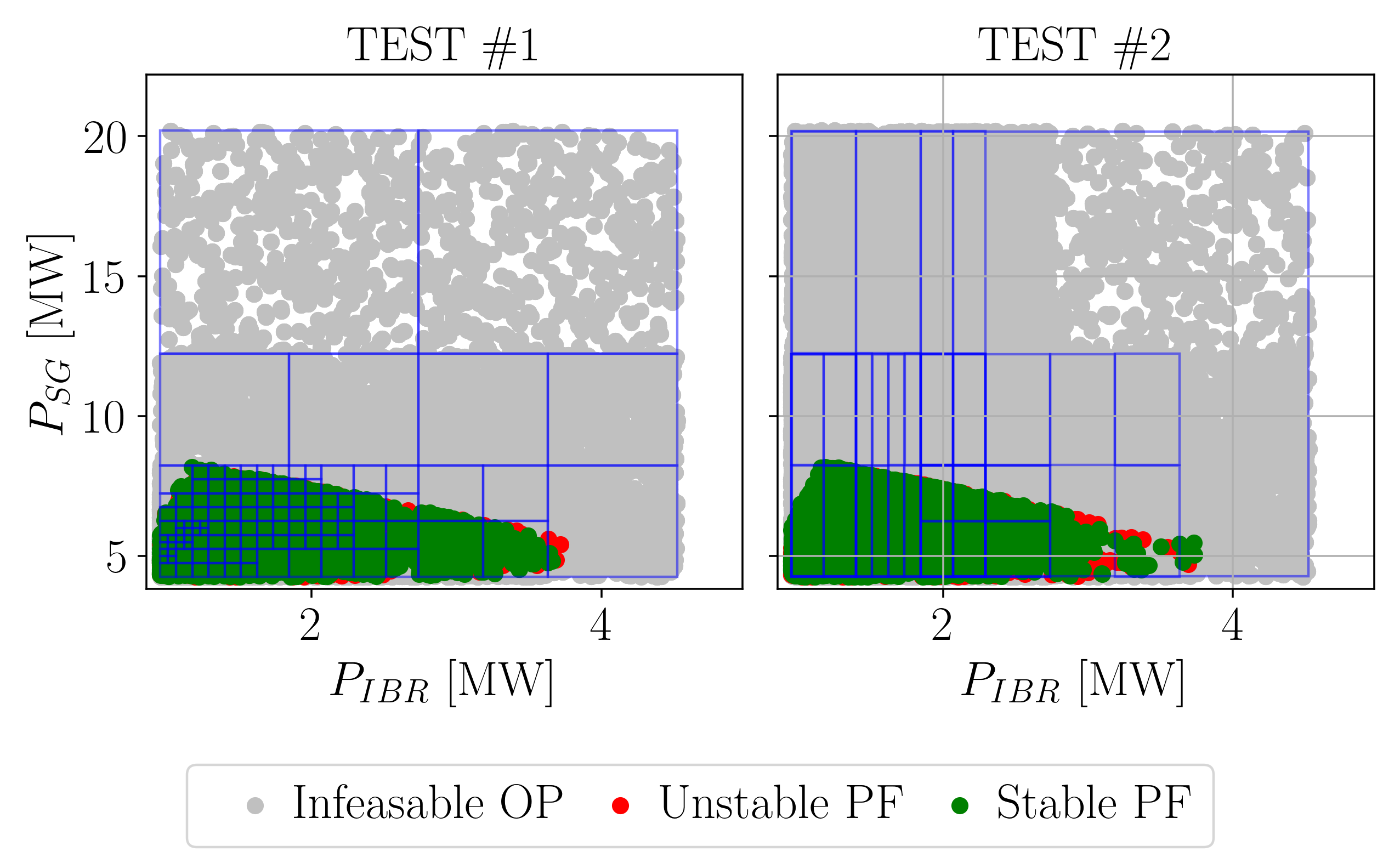}
    \caption{Infeasible \glspl{OP} and stable and unstable \gls{PF} solutions, together with the mesh of subdivisions in the $P_{\mathrm{IBR}}$ and $P_{\mathrm{SG}}$ plane, for the two test cases.}
    \label{fig:p_sg_p_cig_stable_points}
\end{figure}

\begin{figure}
    \centering
    \includegraphics[width=0.7\linewidth]{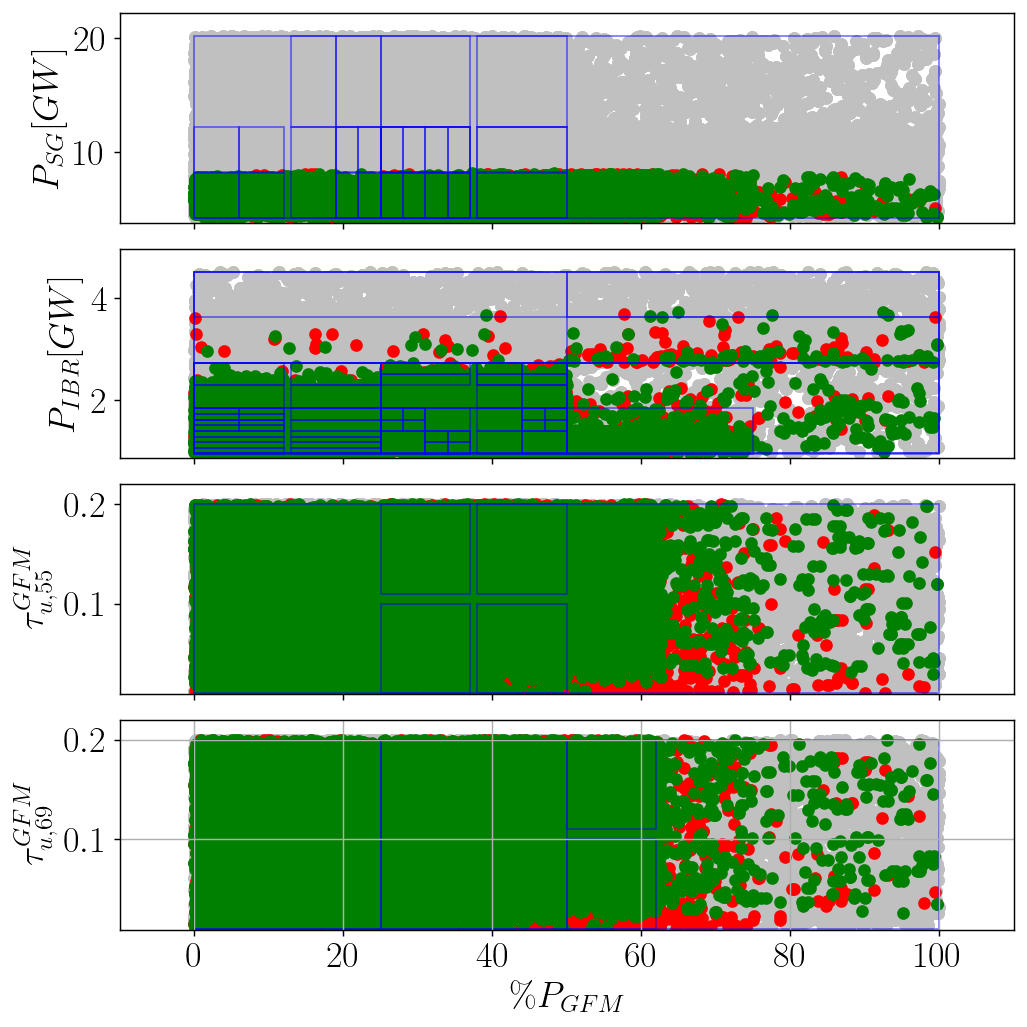}
    \caption{Infeasible \glspl{OP} and stable and unstable \gls{PF} solutions, together with the mesh of subdivisions in the different planes explored as a consequence of the sensitivity analysis applied in Test~\#2.}
    \label{fig:perc_g_for_vars_stable_points}
\end{figure}

\begin{figure}
    \centering
    \includegraphics[width=0.6\linewidth]{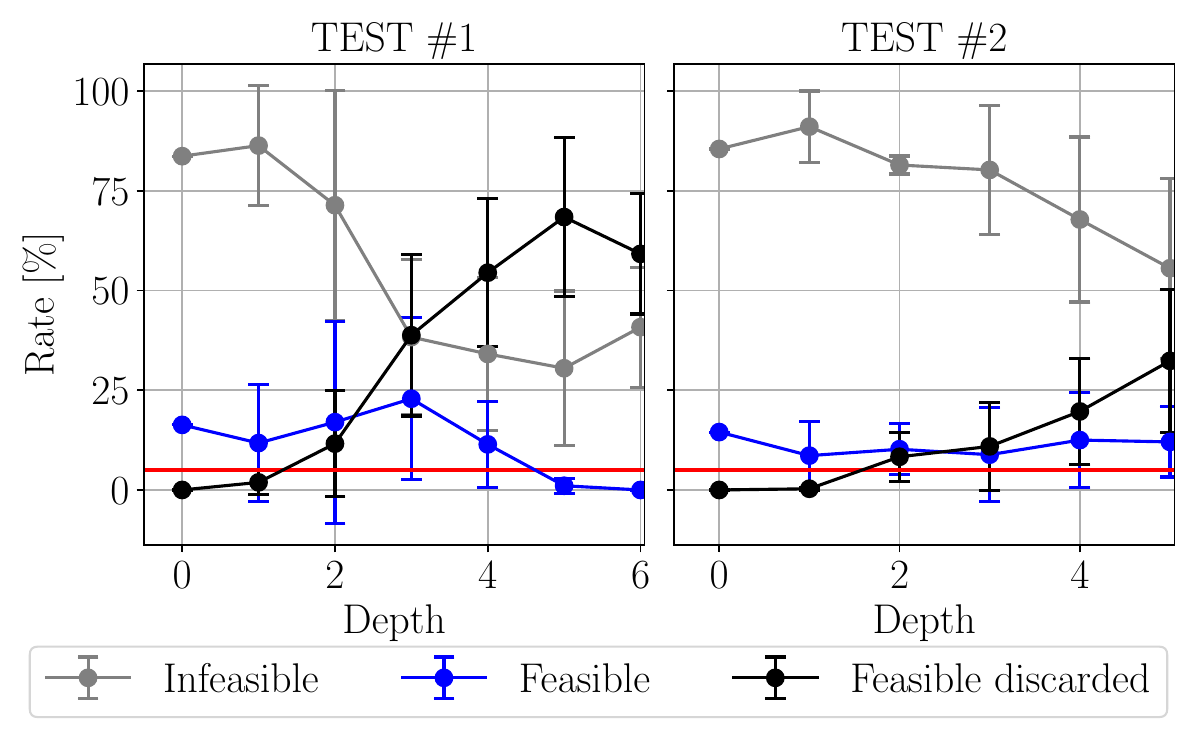}
    \caption{Evolution of feasible, infeasible, and discarded rates as a function of depth.}
    \label{fig:feas_rate}
\end{figure}

Two datasets are created, and their main difference lies in whether sensitivity information is used during the data generation process. In Test \#1, sensitivity is not used; in this case, $P_{SG}$ and $P_{IBR}$ are always split, following the approach in \cite{rossi2022data}. In Test \#2, sensitivity is used to determine along which dimension the split should be performed, selecting the one that most strongly affects the stability outcome before splitting each cell. This aims to demonstrate the necessity of incorporating sensitivity information for larger systems and the scalability improvements enabled by the proposed tool. The other difference in the execution of the data generation process concerns the maximum exploration depth allowed. All parameter settings are summarized in Table~\ref{tab: settings_and_results}. The main characteristics of the resulting datasets (namely, the percentage of feasible, discarded, and stable instances) are also reported.
\begin{table}
\caption{Summary of parameter settings and datasets composition}
\label{tab: settings_and_results}
\begin{tabular}{cccc}
\hline
 &  & TEST \#1 & TEST \#2 \\ \hline
\multirow{5}{*}{Settings} & $n_{samples}$ & \multicolumn{2}{c}{333} \\
 & $n_{cases}$ & \multicolumn{2}{c}{3} \\
 & Max. depth & 10 & 5 \\
 & Min. feasible rate & 5\% & 5\% \\
 & Use sensitivity & False & True \\ \hline
\multirow{7}{*}{Results} & \begin{tabular}[c]{@{}c@{}}Total number\\ of instances\end{tabular} & 152,847 & 164,835 \\
 & Feasible cases & 6.01\% & 11.52\% \\
 & Feasible discarded & 56.82\% & 24.34\% \\
 & Infeasible & 37.17\% & 64.14\% \\
 & Stable instances & 64.54\% & 61.99\% \\
 & Achieved max. depth & 6 & 5 \\
 & Stopping criterion & Min. feasible rate & Max. depth  \\ \hline
\end{tabular}
\end{table}
The percentage of infeasible points is high in both datasets, and particularly high in Test~\#2. Figure~\ref{fig:p_sg_p_cig_feasible_points} shows that a large portion of the operating space in highly loaded conditions is entirely infeasible. Subdividing the initial space along the $P_{\mathrm{SG}}$ and $P_{\mathrm{IBR}}$ dimensions allows these regions to be identified and excluded from sampling. This is illustrated in Figure~\ref{fig:p_sg_p_cig_stable_points}, which displays the operating points and the resulting mesh of splits in the $P_{\mathrm{IBR}}$,$P_{\mathrm{SG}}$ plane. In Test~\#1, the upper regions corresponding to infeasible areas are detected early and are not further split or sampled. In Test~\#2, since splits occur along other dimensions, these regions continue to be sampled, which increases the number of infeasible points. On the other side, always splitting along the same two directions significantly increases the number of points that are discarded because their feasible power flow solutions fall outside the cell being explored. This occurs because the cell dimensions are greatly reduced in Test~\#1, whereas in Test~\#2 the cell dimensions remain larger for almost the same achieved maximum depth (5 in Test~\#1 and 6 in Test~\#2). This effect is visible in Figure~\ref{fig:p_sg_p_cig_stable_points} and Figure~\ref{fig:perc_g_for_vars_stable_points}, where the sampled points are represented in planes associated with other frequently selected splitting dimensions. The results of the sensitivity analysis also allow us to assess the importance of the percentage of \gls{gfor} penetration and of some control tuning parameters for stability. Figure~\ref{fig:feas_rate} shows the evolution of feasible rates as a function of the depth, reporting the mean values and the corresponding standard deviations over the cells at each depth for the infeasible, feasible, and feasible discarded points. The results confirm that in Test~\#1 the infeasible rate decreases with depth, but the rate of feasible discarded points increases substantially, causing the feasible sampling rate to fall below the threshold of 5\%, marked with the red line in Figure~\ref{fig:feas_rate}. Conversely, in Test~\#2, although the infeasible rate is higher, the discarded rate remains low enough to maintain an acceptable feasible sampling rate.

\begin{figure}
    \centering
    \includegraphics[width=0.6\linewidth]{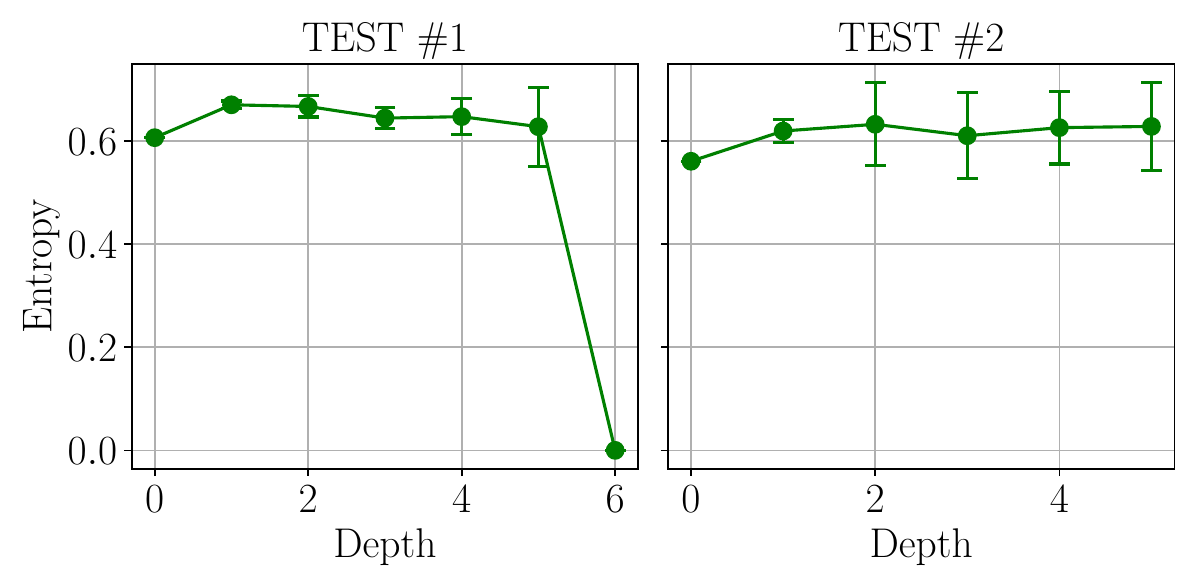}
    \caption{Evolution of entropy as a function of depth.}
    \label{fig:entropy}
\end{figure}

\begin{figure}
    \centering
    \includegraphics[width=0.5\linewidth]{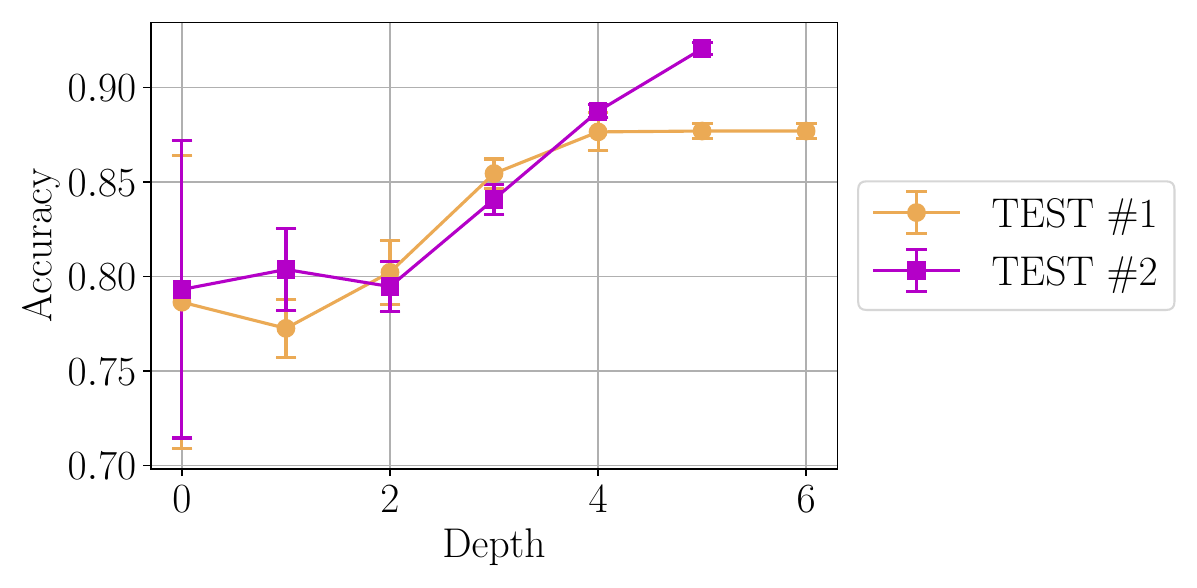}
    \caption{Evolution of \gls{XGB} model accuracy as a function of exploration depth for the two test cases.}
    \label{fig:accuracy}
\end{figure}

Concerning the entropy, which is the other criterion guiding the space exploration both toward the stability margin and as a cutoff condition for stopping the search, Figure~\ref{fig:entropy} shows its evolution as a function of depth. In both test cases the entropy remains above 60\%, except at depth~6 of Test~\#1, where only infeasible and discardable points are found in the explored cells. Note that a perfectly balanced condition corresponds to an entropy of $0.6931$, therefore the entropy-based search strategy ensures a sufficiently balanced dataset.
Finally, an \gls{XGB} model is trained to evaluate the accuracy of the machine learning approach for stability assessment. Figure~\ref{fig:accuracy} shows the evolution of the accuracy as a function of the exploration depth, meaning that the training data used at each point correspond to all samples obtained up to that depth. In this case, the error bars represent the standard deviation of the accuracy, computed using a $k$-fold cross-validation with $k=5$. An increasing trend is observed in both test cases. However, while the model trained with the data from Test~\#1 reaches a plateau, the model trained with the data from Test~\#2 continues to improve. Moreover, the accuracy achieved in Test~\#2 is higher and reaches approximately $92\%$.

\section{Conclusion}
\label{sec: conclusion}
This paper presented a data generation framework for small-signal stability studies of large-scale power systems with high penetration of \glspl{ibr}. The proposed tool addresses key limitations of existing sampling approaches by combining scalable operating-space design, adaptive exploration guided by online sensitivity analysis and entropy-based criteria, and full integration with \gls{HPC} environments.

The case study demonstrated the applicability and benefits of the proposed framework. The sensitivity-driven sampling strategy produced a dataset with higher feasible-sample density and superior performance when used to train machine learning models for stability assessment. In contrast, the non-sensitivity approach generated a larger proportion of infeasible and discarded points, confirming that sensitivity-guided exploration is essential for scalable data generation in large and complex systems.

The data generation framework developed in this work is released as an open-source tool, fully implemented in Python and publicly available on GitHub at \url{https://github.com/MauroGarciaLorenzo/datagen}, with the aim of facilitating reproducibility, encouraging collaboration, and supporting further developments in scalable stability assessment and data-driven applications for power systems with high levels of \glspl{ibr}.

\appendix
\section{Dynamic Equivalents of IBRs Aggregation}
\label{appendix: dyn_eq}
\begin{figure}
    \centering
    \includegraphics[width=0.5\linewidth]{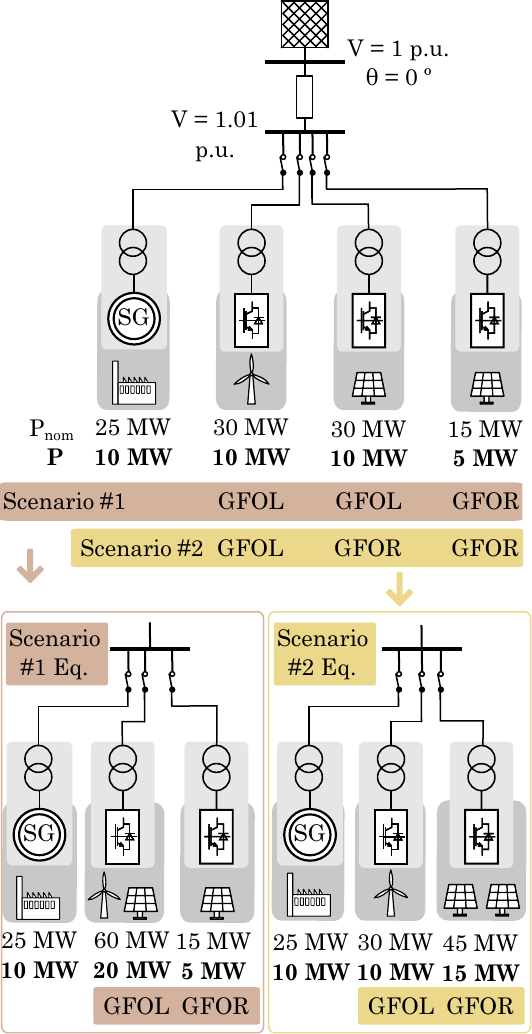}
    \caption{Two operating scenarios and their equivalent, in terms of \gls{gfor} or \gls{gfol} converters aggregation.}
    \label{fig:scenarios_example}
\end{figure}

\begin{figure}[htbp]
    \centering

    \subfigure[Scenario \#1 and its equivalent.\label{fig:Yscan_scen1}]{
        \centering
        \includegraphics[width=0.48\linewidth]{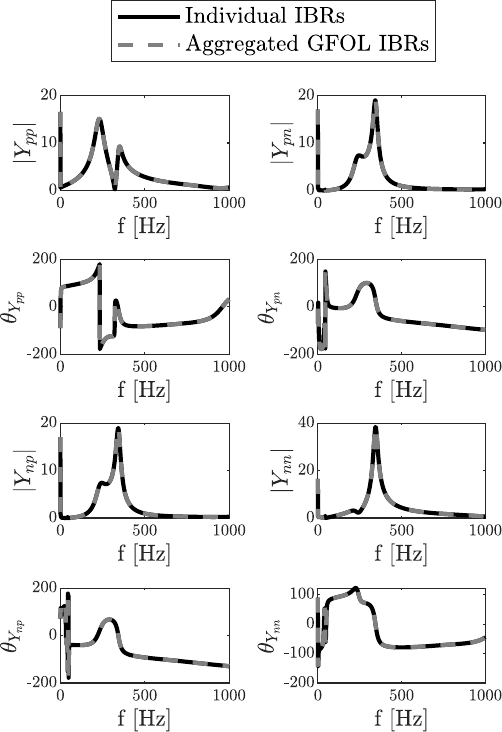}
        }
    \hfil
    \subfigure[Scenario \#2 and its equivalent.\label{fig:Yscan_scen2}]{
        \centering
        \includegraphics[width=0.48\linewidth]{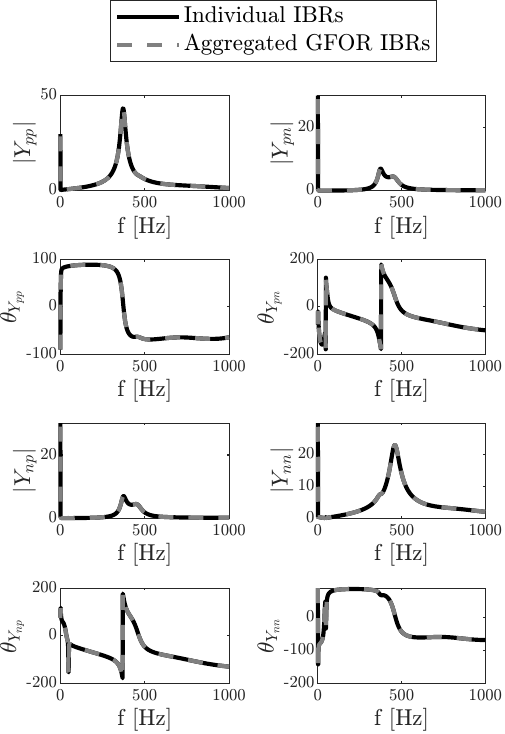}
        }

    \caption{Admittance frequency-scan comparison.}
    \label{fig:mainfig}
\end{figure}

This section presents the admittance-based frequency scan analysis carried out to verify the dynamic equivalence of the \glspl{ibr} aggregation employed in the data generation tool. The comparison between the admittance-based frequency scans of the base scenarios (with individually modeled \glspl{ibr}) and the corresponding aggregated scenarios is performed under consistent operating conditions. The aggregated converter is assigned a nominal power equal to the sum of the nominal powers of the individual converters it represents, and its operating point corresponds to the total active power output of those converters. Additionally, the voltage at the bus to which the \glspl{ibr} are connected, as well as the voltage at the bus representing the Thevenin equivalent of the remainder of the grid, are kept equal across both scenarios.
Figure~\ref{fig:scenarios_example} illustrates two base scenarios along with their corresponding aggregated equivalents. Figures~\ref{fig:Yscan_scen1} and~\ref{fig:Yscan_scen2} present the comparison between the respective admittance-based frequency scans, demonstrating a perfect match between the responses obtained in the base scenarios and those derived from the equivalent aggregated representations.

\section{Tools for Power System Operation and Stability Analysis}
\label{appendix: SS_FEASPF}
To assess the small-signal stability of the sampled operating points and associated ~\glspl{ibr} control settings, conventional power system stability analysis tools are employed. These tools typically rely on the linearized state-space representation of the system. The eigenvalue analysis of the system’s state matrix provides insight into system stability. 
The linearization process is performed around an equilibrium point, which can be obtained either through time-domain simulations or \gls{PF} calculations. In this work, the latter approach is preferred due to its computational efficiency. It is important to note that, despite the sampling strategy being designed to respect key relationships, 
it is still likely that many sampled operating points do not satisfy the system's feasibility constraints. This is primarily due to the voltage profiles that, even if sampled considering the existing relation between adjacent buses, may lead to unrealistic or infeasible steady-state conditions, such as reactive power injections from generators that violate their capability curves. Additional sources of infeasibility include violations of line thermal ratings, voltage magnitude limits, and generators power factor. To ensure that only feasible operating points are analyzed, these constraints must be accounted for. Therefore, instead of performing a standard \gls{PF} calculation, an \gls{OPF} problem is solved. The objective function minimizes the squared difference between the sampled active power set point of each generator and its resulting active power injection. 
This formulation allows the system to adjust controllable variables within operational limits to satisfy all relevant constraints, ensuring that the resulting operating point is both physically realistic and technically feasible. To implement these calculations in a Python environment, the feasible operating point is computed using the VeraGrid~\cite{veragrid} library. The AC-\gls{OPF} implemented in VeraGrid, which considers voltage, current, and generator power factor constraints as described in~\cite{alegre2024integration}, has been adapted to incorporate the objective function described above. For the small-signal stability analysis, the Python version of the STAMP tool~\cite{arevalo2025matlab} is used, which is available at~\cite{stampy}.


\section*{Declaration of Generative AI and AI-assisted technologies in the writing process}
During the preparation of this work, the authors used the tool ChatGPT developed by OpenAI to improve the readability. After using this tool/service, the authors reviewed and edited the content as needed and take full responsibility for the publication’s content.

\section*{Acknowledgments}
This work was supported by the Project TED2021-130351B-C21 (HP2C-DT), funded by MICIU/AEI/10.13039/501100011033 and by the European Union NextGenerationEU/PRTR.






\end{document}